\documentclass[10pt,twocolumn]{article}

\usepackage{color}
\usepackage{graphicx}
\usepackage{multicol}
\usepackage{verbatim}
\usepackage{authblk}

\title{Seasonal footprints on ecological time series and jumps in dynamic states of protein configurations from a nonlinear forecasting method characterization}
\author[1,*]{Leonardo Reyes} 
\affil[1]{{\small Laboratorio de Din\'amica no-lineal y Sistemas Complejos, Centro de F\'isica, Instituto Venezolano de Investigaciones Cient\'ificas (IVIC), Rep\'ublica Bolivariana de Venezuela.}}
\author[2]{Kilver Campos}
\author[2]{Gilberto D. Avenda\~no}
\affil[2]{{\small Laboratorio de F\'isica de Fluidos y Plasmas, Centro de F\'isica, Instituto Venezolano de Investigaciones Cient\'ificas (IVIC), Rep\'ublica Bolivariana de Venezuela.}}
\author[3]{Lenin Gonz\'alez-Paz}
\author[3]{Alejandro Vivas}
\affil[3]{{\small Laboratorio de Biocomputaci\'on, Centro de Biomedicina Molecular, Instituto Venezolano de Investigaciones Cient\'ificas (IVIC), Maracaibo, Edo Zulia, Rep\'ublica Bolivariana de Venezuela.}}
\author[4]{Ysa\'ias J. Alvarado}
\affil[4]{{\small Laboratorio de Qu\'imica Biof\'isica Te\'orica y Experimental, Centro de Biomedicina Molecular, Instituto Venezolano de Investigaciones Cient\'ificas (IVIC), Maracaibo, Edo Zulia, Rep\'ublica Bolivariana de Venezuela.}}
\author{Sa\'ul Flores}
\affil[5]{{\small Laboratorio Ecolog\'ia de Suelos, Centro de Ecolog\'ia, Instituto Venezolano de Investigaciones Cient\'ificas (IVIC), Rep\'ublica Bolivariana de Venezuela.}}
\affil[*]{{\small corresponding author: leonardoivanrc@gmail.com}}

\begin{document}

\maketitle

\begin{abstract}
We have analyzed phenology data and protein configurations from molecular dynamics simulations with the nonlinear forecasting method proposed by May and Sugihara. Our primary focus in this work is to characterize the dynamic state of a system by quantifying prediction quality from data. Full plots of prediction quality as a function of dimensionality $E$ and forecasting time $T_p$, the two basic parameters of the method, give fast and valuable information about Complex Systems dynamics. We detect changes in protein dynamics due to mutations and, regarding ecology data, we show how cycles and {\it rhythms} of the environment manifests in parameter space $(E,T_p)$ for some species.\\
\end{abstract}

\begin{figure}[!ht]
\centering
\includegraphics[width=0.5\textwidth]{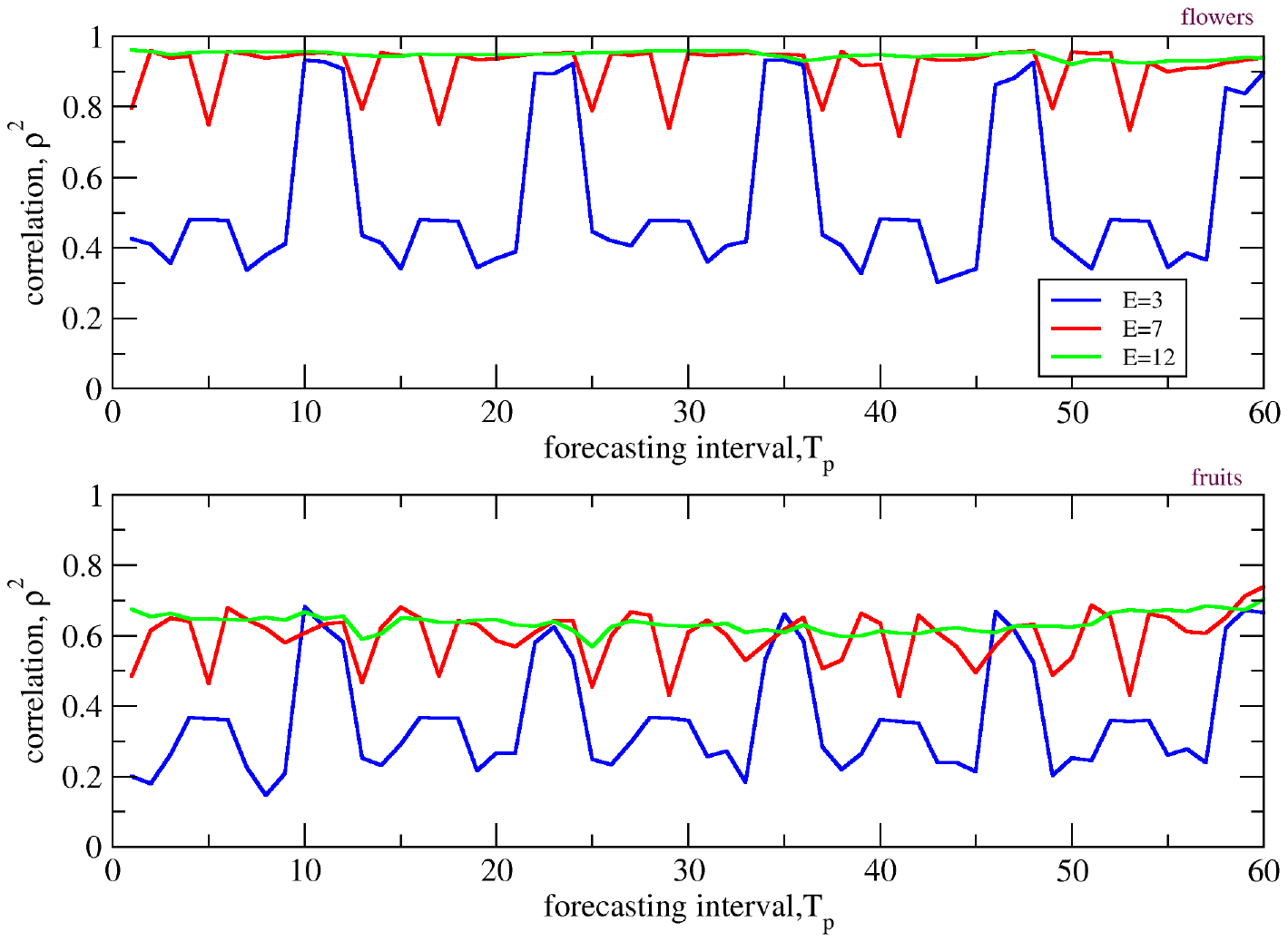}
\caption{$\rho^2(T_p)$ for $E=3$, $E=7$ and $E=12$ ({\it Hieronyma moritziana}), flowers on top and fruits at the bottom. For $E=3$ we get seasonal scales from the signal. 
$T_p$ is given in months. $E=7$ is the first case outside the {\it small dimensionality} region in parameter space for this species. We consider only the first $350$ data points, see figure \ref{windows} and text.}
\label{hyronima_rho_vs_tp}
\end{figure}

The notion of {\it prediction} is at the heart of the scientific method \cite{MS90} and of some theories about the mind (the predictive brain) \cite{brain1}. 
In this article we explore decades of ecological data \cite{EcologySaul1} and data from a model for sampling protein configurations using a dynamic indicator $\rho(E,T_p)$ introduced some time ago by May and Sugihara to distinguishing chaos from noise in time series \cite{MS90} : $\rho^2$ is a number in the unit interval which {\it quantify the quality of the predictions} made by the forecasting method introduced by the authors.
This is a method that requires neither large ammounts of data nor preprocessing of raw data. Besides, no previous training is needed as is usual in
machine learning. From our implementation of the method written in the C programming language 
we have obtained fast and valuable insights about such data when considering full plots of the quantity $\rho^2(E,T_p)$. 
We will argue that this can be used as a general dynamic characterization of a system which also allow us to consider in a novel way a framework 
for {\it adaptation} and {\it prediction} in Complex Systems.

\begin{figure}[!ht]
\centering
\includegraphics[width=0.4\textwidth]{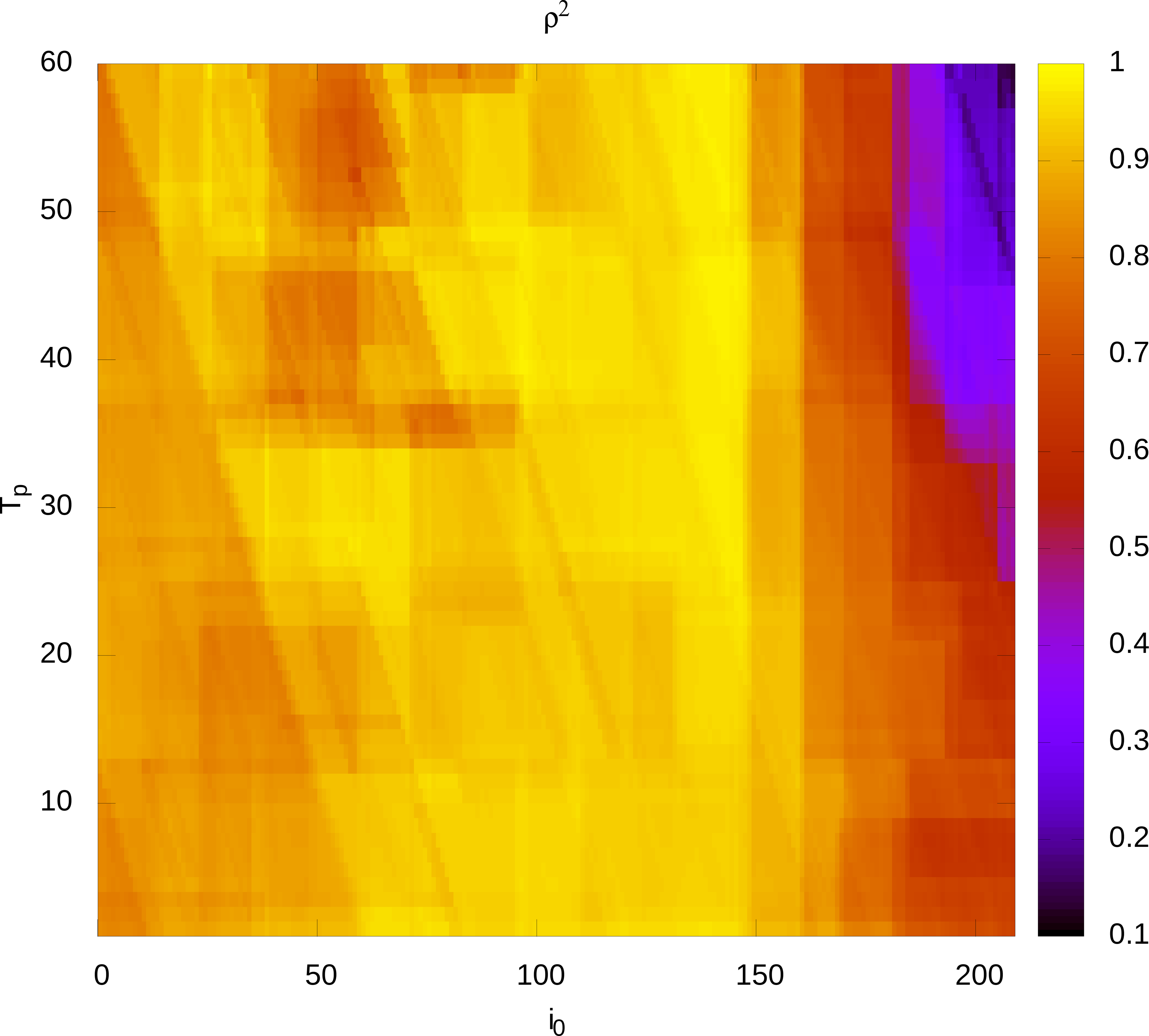}
\caption{For {\it Hieronyma moritziana} (flowers) we show $\rho(i_0,T_p)$ for high dimensionality ($E=12$), 
for windows of data size $M=N/2$ that starts at $i_0$ (see text). Here $N$ is the number of data points ($N=420$ for the data shown).}
\label{windows}
\end{figure}

Here $E$ is a dimensionality parameter: 
in the forecasting method is the number of components of vectors (see below) constructed with the data. 
We make a prediction $T_p$ {\it time} steps into the {\it future}
and the evaluation of that prediction is made with the quantity $\rho$ by comparing predictions with actual observed values.
Characterizing the dynamics of a system is a basic goal in today's big data dominated scientific practice.
In this work we have applied the method proposed by May and Sugihara to analyze data spanning decades of population dynamics of several species from a Venezuelan forest \cite{EcologySaul1} and to analyze some aspects of protein dynamics \cite{protein1}.
We have observed {\it chaotic}, {\it disarmed} and {\it rigid} transient states in protein configurations. Also the method clearly detects the
effect of mutations in the collective dynamics of a typical intrinsically disordered protein such as alpha synuclein, a protein that is associated with neurodegenerative Parkinson's disease. For ecological data from a Venezuelan forest we have observed a qualitative change in the dynamics
as measured by $\rho(E,T_p)$. Previous to this transition the dynamics of certain species can be modeled with a random decay process with reinsertion.
Besides, the method allow us to quantify how sensitive is a particular species to the environment's cycles and {\it rhythms}.
By design, we lose reference to any particular {\it time} in phenology data or {\it position} in protein data, a feature that can be 
overcome by considering blocks or windows of data (see fig. \ref{windows}).

Given some data series $x_n$ the nonlinear forecasting model considers vectors of $E$ components of the form $(x_i,x_{i-1},x_{i-2},\ldots,x_{i-(E-1)})$ and
constructs a prediction for the next $x_{i+T_p}$ value in the series; the prediction's quality is quantified with the correlation $\rho$ between observed and predicted values. Thus, for given data $x_n$, the quantity $\rho(E,T_p)$ is the basic output of the method \cite{MS90}. 
Specifically, for each vector of $E$ components we look for the $E+1$ more similar vectors in the data and a prediction is made by a weighted average of the observed $T_p$ times into the future values for each of the $E+1$ more similar vectors. The weight is chosen as the exponential of the euclidean distance between vectors: the closer to the original vector of $E$ components the larger the weight. We then quantify the quality of our prediction with the correlation coefficient $\rho$ between observed and predicted values: if the dynamics is deterministic then similar past values should produce similar future values.
As in \cite{MS90} we consider every possible vector of $E$ components from the second half of the data and look for the more similar ones in the first
half of the data.

In figure \ref{hyronima_rho_vs_tp} we show $\rho^2(T_p)$ for {\it Hieronyma moritziana Pax \& K. Hoffm} species for three values of the dimensionality parameter $E$, $E=3$, $E=7$ and $E=12$, 
for signals of flowers and fruits populations as a function of time \cite{EcologySaul1} (we used only the first $350$ data points, see below and figure \ref{windows}).
It can be appreciated that the seasonal time scale (one year) manifests itself at $E=3$, where we obtain a periodic pattern. 
For high dimensionality ($E=12$) we obtain that $\rho$ is weakly dependent on $T_p$.

\begin{figure}[!ht]
\centering
\includegraphics[width=0.45\textwidth]{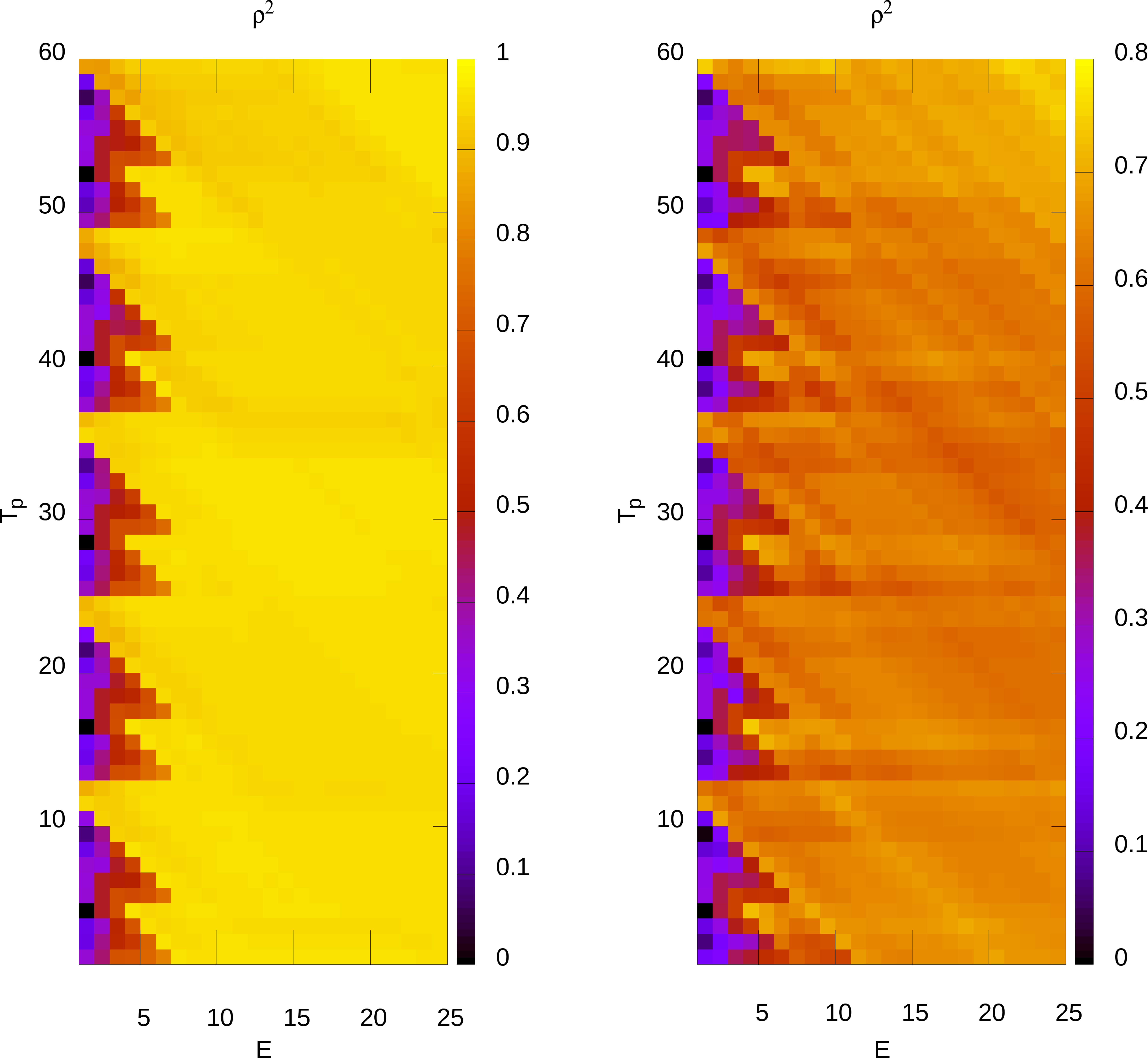}
\caption{$\rho^2(E,T_p)$ for {\it Hieronyma moritziana} species: flowers (left) and fruits (right). We consider the first $350$ data points, see figure \ref{windows} and text.}
\label{hyronima_scan}
\end{figure}

\begin{figure}[!ht]
\centering
\includegraphics[width=0.45\textwidth]{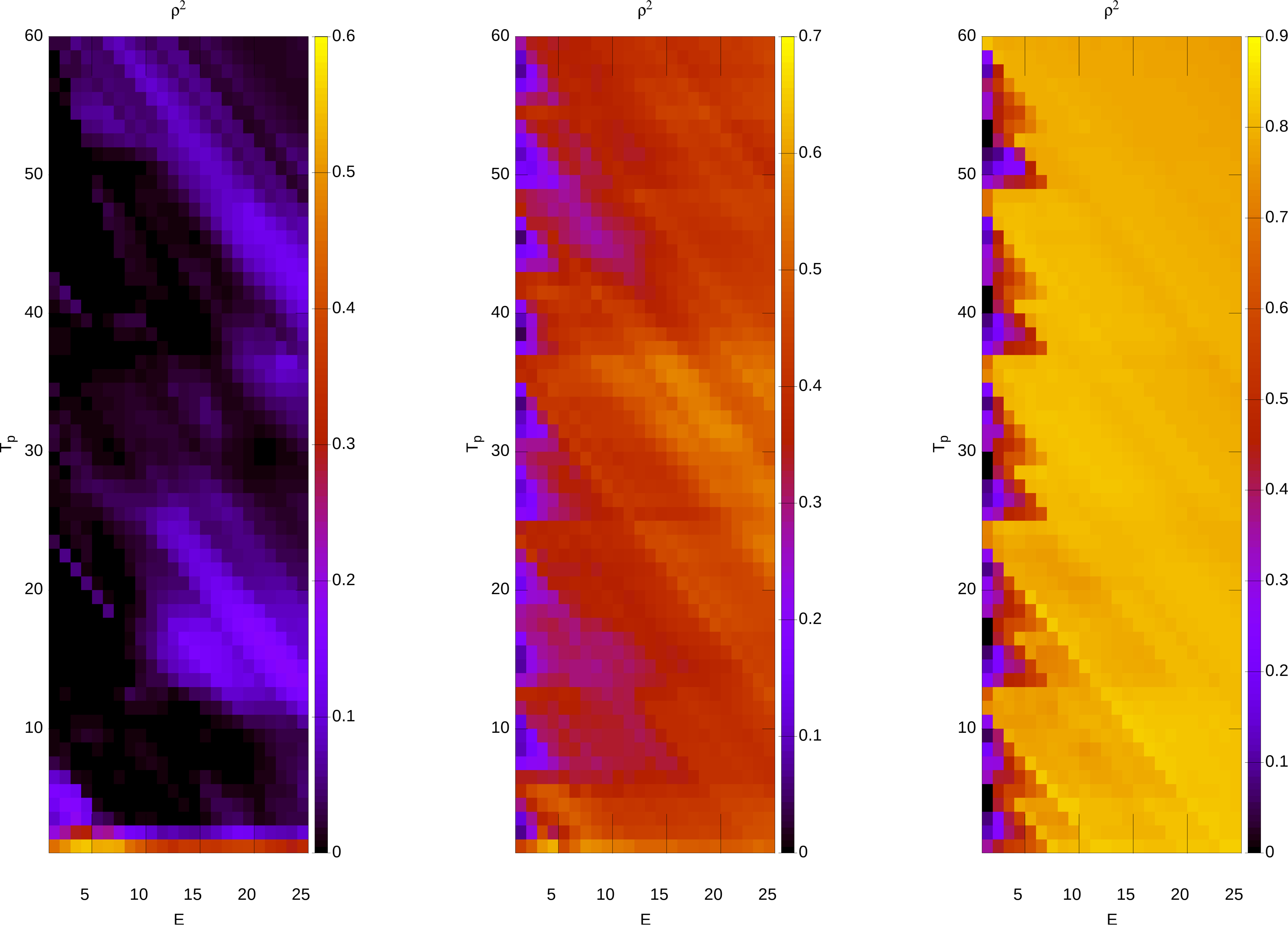}
\caption{Three cases of population dynamics (flowers) for species with very different predictability patterns in parameter space. {\it left}: Richeria Grandis, {\it center}: Palicurea Fendleri, {\it right}: Caracasia Tremadena. See also \cite{EcologySaul1}.}
\label{figtriple}
\end{figure}

In figure \ref{windows} we check for stationarity in the phenology signals. 
We consider windows of data of size $M$ that start at $i_0$. For example, to consider the first half of the data 
is to consider $M=N/2$ and $i_0=0$, to consider the second half of the data is to consider $M=N/2$ and $i_0=N/2$.
It can be observed that there was a qualitative change in the species's dynamics as measured by $\rho^2$, so in figures \ref{hyronima_rho_vs_tp}, 
\ref{hyronima_scan} and \ref{figtriple} we used only the first $350$ data points in order to consider the first stage of the dynamics. By the end of the data, for large $i_0$, we have a high dimensionality region in which $\rho$ decays with $T_p$. An abrupt change appears for $i_0\sim 150$ in figure \ref{windows},
the same occurs for small dimensionality and other species, this suggest an ecosystemic collapse. The smaller $M$ the best localization of the transition point, in principle, but for smaller $M$ he have also poorer statistics for $\rho$. See below for more on the late stage.

Full plots of $\rho(E,T_p)$ can be very informative about the dynamic state of a system. 
In figure \ref{hyronima_scan} we show $\rho(E,T_p)$ for {\it Hieronyma moritziana} species, and some features emerge:
only for $E<7$ the seasonal scale appears, and there are triangular patterns that repeat themselves each $12$ months. 
For flowers and $E>6$ we obtain that $\rho$ is weakly dependent on $T_p$ and $E$. 
Other species have no structure in parameter space or some distorted triangular patterns, see figure \ref{figtriple} where we show $\rho^2(E,T_p)$ for three species.

\begin{figure}[!ht]
\centering
\includegraphics[width=0.4\textwidth]{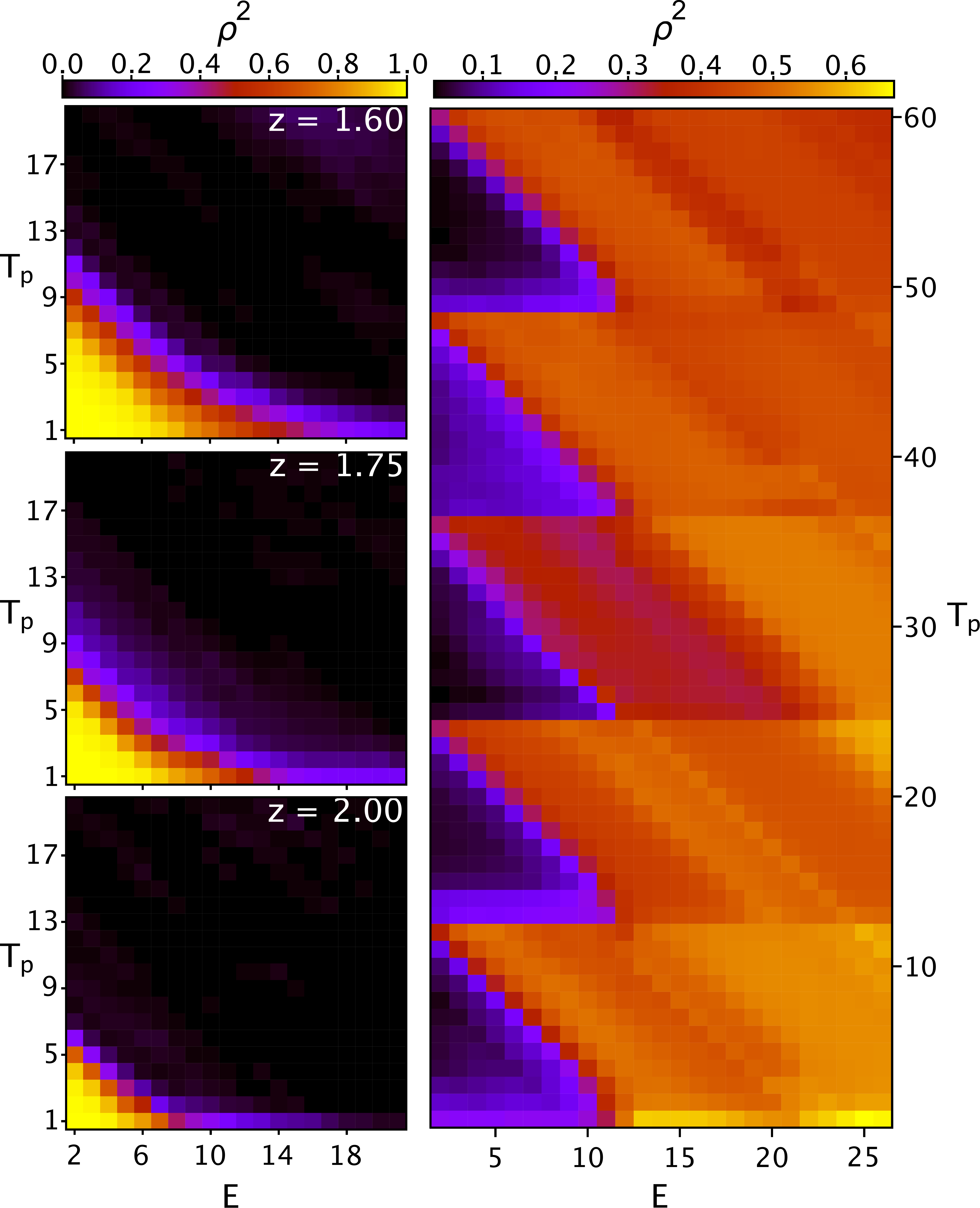}
\caption{{\it left}: $\rho^2(E,T_p)$ for chaotic dynamics with different Lyapunov exponents $\lambda$. From top to bottom: $z=1.6,\ 1.75,\ 2.0$, with $\lambda=\log z$ \cite{Gonzalez1}. 
Similar results were obtained for the logarithmic map. {\it right}: $\rho^2(E,T_p)$ for synthetic data emulating a random decay process with reinsertion. 
We have used $z=2/3$ in the explicit functions of references \cite{Gonzalez1} for the random decay process. We used $500$ data points in each case. See text.}
\label{fig34}
\end{figure}

For comparison and reference we show in figure \ref{fig34} (left) $\rho^2(E,T_p)$ plots for chaotic time series with different Lyapunov exponents \cite{Gonzalez1}.
It can be appreciated that the intersection points on the $E$ and $T_p$ 
axes are the same, and that the larger the Lyapunov exponent the smaller those intersection points. 
Similar triangular patterns were obtained for {\it Hieronyma moritziana} like species but with a major difference (see figure \ref{hyronima_scan}):
for several species we have obtained a kind of inversion and repetition of the triangular patterns obtained for chaotic series, which prompts the question:
how can we obtain low prediction quality at short times and high prediction quality at longer times?. 
We have found that a random decay process with resinsertion \cite{Schuster} produces
the same kind of triangular patterns obtained for {\it Hieronyma moritziana} like species, 
this result is shown in figure \ref{fig34} (right). We have generated a random decay signal with the explicit functions
of Gonz\'alez {\it et al}: $X_n=\sin^2(\theta\pi z^n)$, with parameter $z=p/q$, $p$ and $q$ being relative primes ($q>p$, $z<1$) \cite{Gonzalez1}. 
A random decay with reinsertion sounds very {\it ecosystemic} indeed. We get poor prediction at short times because it's a random process, and we obtain a high prediction
quality at longer times because once the signal has decayed to near zero values the dynamic is very predictable.
It can be observed in figure \ref{hyronima_scan} that for {\it Hieronyma moritziana} and within one year we obtain a kind of superposition of two triangles (flowers mainly), 
corresponding to two overlapping random decay processess: wet and dry periods of the year in the Venezuelan forest \cite{EcologySaul1}. 
A simpler plot like the one shown in figure \ref{fig34} (right) highlights the underlying proposed mechanism.
Since we have random decay processess of size $6$, it can be said that the case $E=7$ (see figure \ref{hyronima_rho_vs_tp}) is the smaller dimensionality value $E$ outside the {\it small dimensionality} region in parameter space for {\it Hieronyma moritziana}.

\begin{figure}[!ht]
\centering
\includegraphics[width=0.3\textwidth]{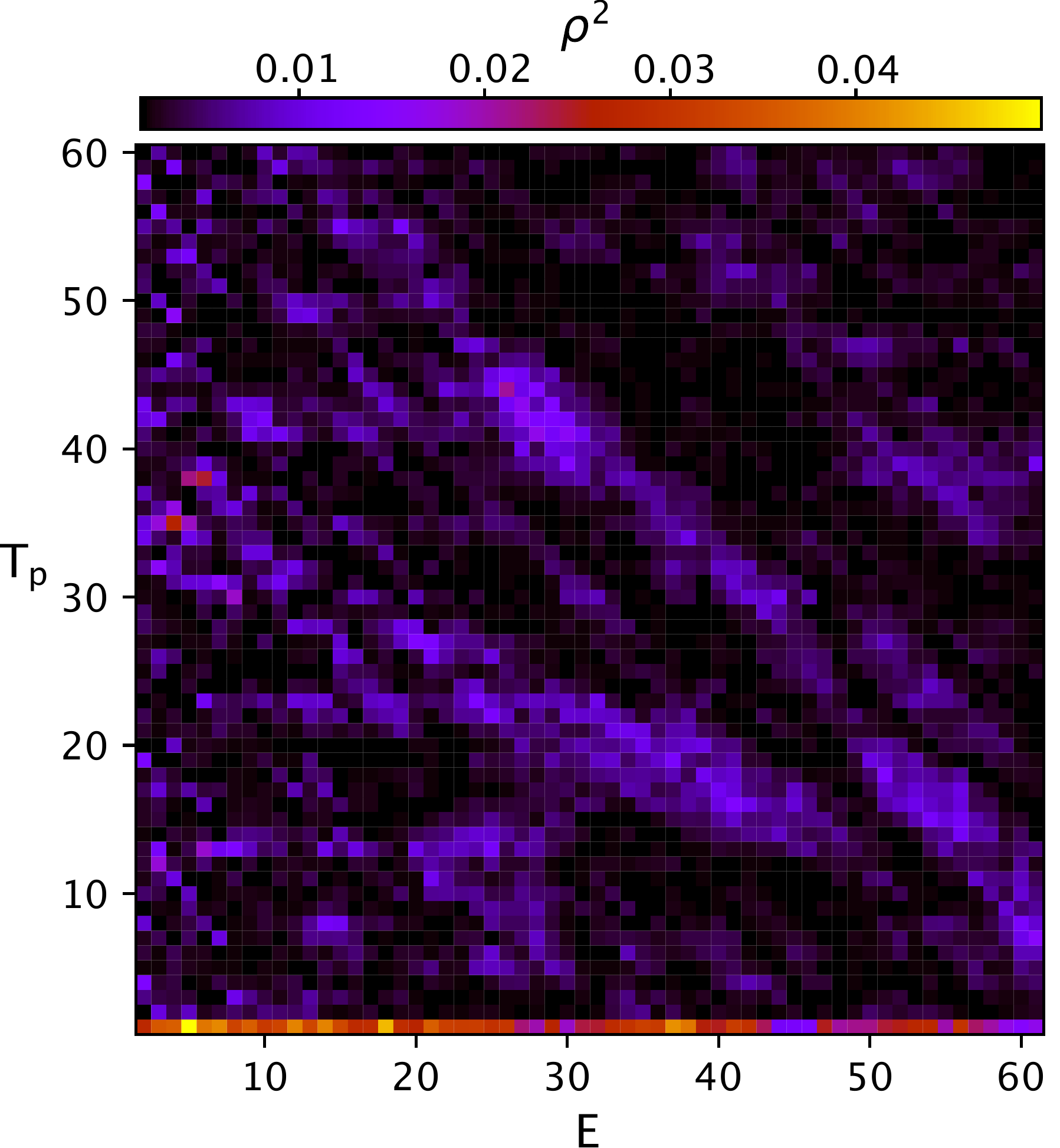}
\caption{$\rho^2(E,T_p)$ for an ideal gas like dynamics. The processed signal was the {\it activity} in the system as a function of time, generated from the GHWS model \cite{ghws} 
in a zone of parameter space that produces zero average correlations of activity (see \cite{Chandler}, section $3.6$), 
which results in the ideal gas equation of state \cite{new}. }
\label{ideal}
\end{figure}

In figure \ref{ideal} we have verified, with the help of the GHWS model \cite{ghws}, that an ideal gas type of dynamics would produce very small values of $\rho^2$, for most $E$ and $T_p$ values. 
The signal used in this case was the {\it activity} in the system as a function of time \cite{ghws,new,Chandler}.

In figure \ref{windows} we showed that there was a qualitative change in the Hieronyma moritziana species dynamics (this species is one of the more sensitive).
Early in the data the high dimensionality region virtualy does not depends neither on $E$ nor $T_p$, appart from fluctuations. This is a zone characterized by a random decay process plus a random reijection. We think that the ``random" part of the ``reinjection process" proposed before should be relaxed in order to explain the new dynamic scenario. 

We have found that what defines the frontier between small and large dimensionality is associatted 
to the ``size" of the random decay process ($E=6$) and also to an {\it overfitting} scenario: an overfitting could be the cause that the one year time scale
does not appears for $E>6$.

\begin{figure}[!ht]
\centering
\includegraphics[width=0.45\textwidth]{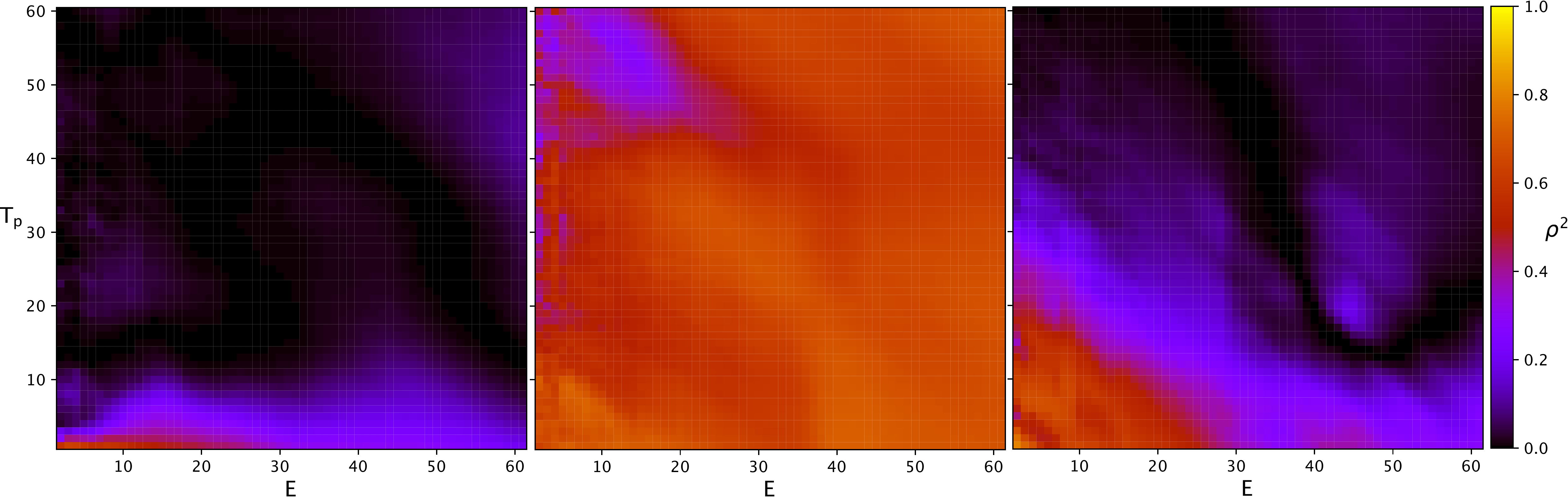}
\caption{Protein {\it dynamics}, for data obtained with molecular dynamics simulations \cite{protein1,protein2,moldyn}. On the left a dynamic state, at a particular time, for which $\rho^2\sim 0$ for most $E$ and $T_p$: the protein is {\it disarmed}. At the center a dynamic state, at a particular time, for which $\rho^2$ is large for most $E$ and $T_p$: the protein is {\it rigid}. To the right 
we show a chaotic like structure obtained for a particular time. }
\label{figprot01}
\end{figure}

\begin{figure}[!ht]
\centering
\includegraphics[width=0.4\textwidth]{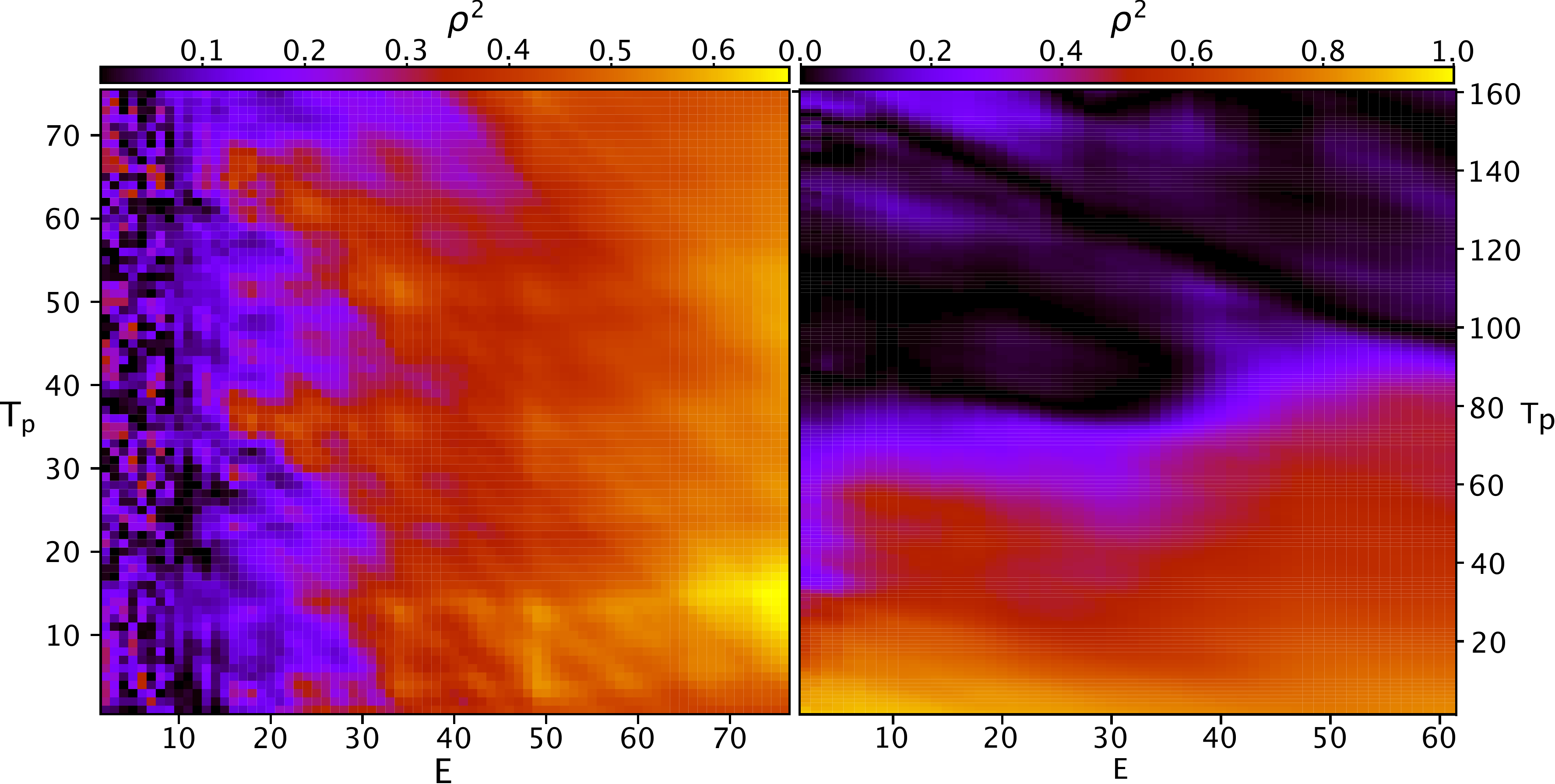}
\caption{{\it left}: $\rho^2(E,T_p)$ from protein dynamics, a particular dynamic state for data obtained with molecular dynamics simulations: $\rho^2\sim 0$ for $E=E_0\sim 12$, for any $T_p$ considered. {\it right}: $\rho^2(E,T_p)$ for $\alpha$-sinuclein protein with mutation A53T, after $100ns$. The input signal was the $y$ coordinate of the atoms in the N domain.}
\label{proteindoble}
\end{figure}

In figures \ref{figprot01} and \ref{proteindoble} we show a $\rho^2(E,T_p)$ plot for the atoms coordinates of a protein 
obtained with molecular dynamics simulations \cite{protein1,protein2,moldyn}. 
We used the raw data: the coordinates (at a particular time) of the atoms in the protein.
As an example, we have chosen a typical intrinsically disordered protein such as alpha synuclein, associated with neurodegenerative Parkinson's disease, which is known for its complex dynamic conformational behavior and tendency to form aggregates \cite{alpha}.
This protein is made up of three domains or regions known as the N-terminal, NAC, and C-terminal domains \cite{alpha2}.
The signal to be processed in this case is the position of the atoms in one of the three regions of the alpha synuclein protein, for a given time.
As time passes towards equilibrium the protein jumps between very different dynamic states (see figure \ref{figprot01}): fully {\it disarmed} ($\rho\sim 0$ for most $E$ and $T_p$), 
{\it rigid} (high values of $\rho$ for most $E$ and $T_p$), a chaotic-like structure (triangular pattern), and several more complicated $\rho(E,T_p)$ patterns.
It is important to note that structural disorder has been highlighted recently for alpha synuclein wild type \cite{alpha3}, and that we
are associating predictability (large values of $\rho^2$) with rigidity through lost of effective degrees of freedom.
A particular pattern emerges for an alpha synuclein dimer scenario: we obtain a particular dimension value $E_0$ for which 
no prediction can be made about the position of the atoms along the polymer chain, for any value of $T_p$, see figure \ref{proteindoble} (left).
Regarding $E_0$, we have interpreted this result as implying that there are structures of lenght $E_0$, involving $E_0$ number of atoms, that appear in the protein (for a given time). 

So far we have shown transient protein dynamics states. 
In figure \ref{proteindoble} (right) we show $\rho^2(E,T_p)$ for the N region of a mutated protein (A53T) after $100ns$ of simulation. 
The N-terminal region of the alpha synuclein protein was chosen given its important biological function of binding to the cell membrane \cite{alphaN,alpha3}.
No such state were achieved for the same protein in the wild type case, and this particular dynamic state, a near equilibrium one, 
fits a chaotic structure scenario: good predictability for small $T_p$ and eventually unpredictability for large $T_p$, but with an approximately horizontal frontier between
large and near zero values of $\rho^2$. These results will be further investigated elsewhere, but let's consider the case of an horizontal frontier (in parameter space) between high and low values
of $\rho^2$. If we look at figures \ref{hyronima_rho_vs_tp} and \ref{hyronima_scan}, we can see that when $T_p$ is a multiple of $12$ months $\rho^2$ shows a weak dependence on dimensionality $E$,
which in turn implies that variations of $\rho$ in parameter space (see below) points only in the $T_p$ direction. This suggest that an approximately horizontal frontier like 
the one shown in figure \ref{proteindoble} (right) could be associated with an interaction with large parts of the protein (the equivalent of {\it seasonal} scales in this case).
Under these considerations we can say that the mutated protein gains some (pathological) rigidity because of self-interaction.
Besides, it appears that this pathology could be associated to a lack of cost to make predictions, since the dimensionality parameter $E$ 
can be directly related to memory capacity which certainly it is expected to have a {\it cost}.

\begin{figure}[!ht]
\centering
\includegraphics[width=0.45\textwidth]{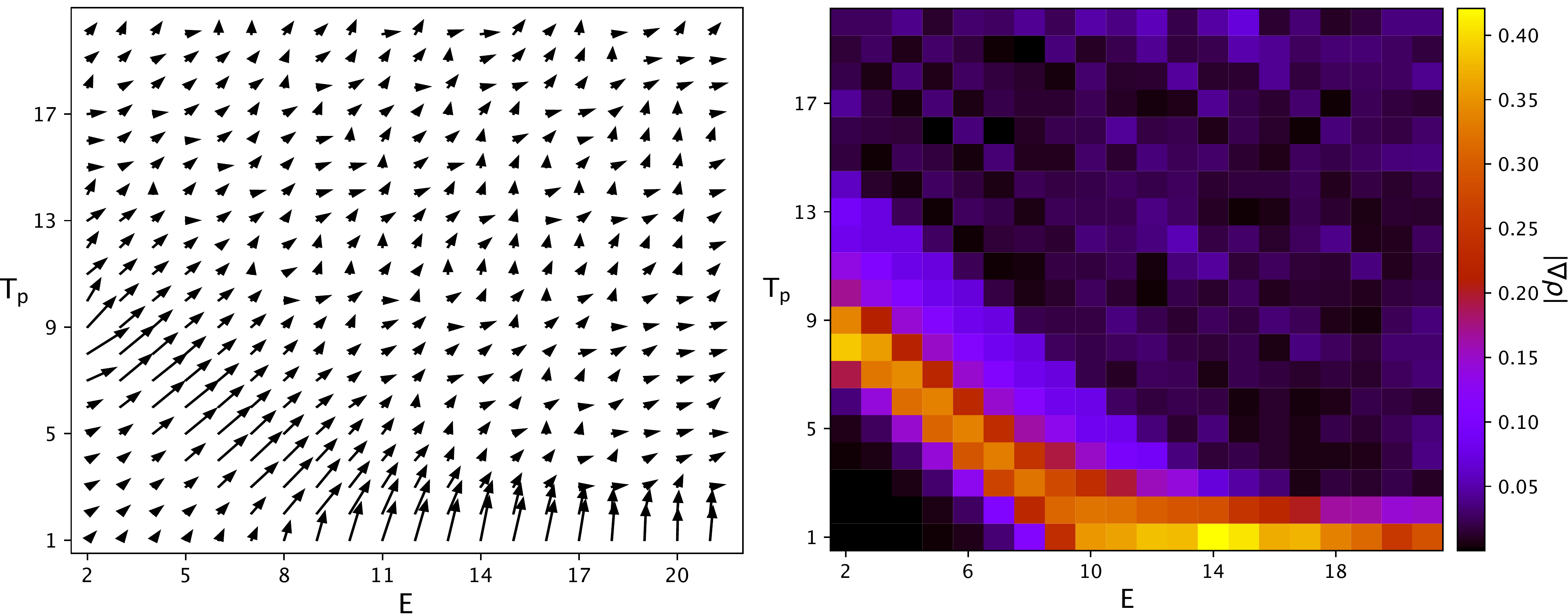}
\caption{For data obtained from the logistic map (full chaos) we show the vector field $\nabla\rho$ (left) and the modulus of the gradient of $\rho$ in parameter space $|\nabla\rho|$ (right). 
We used $500$ data points. See text.}
\label{figGrad}
\end{figure}

{\it Complex adaptive systems}. If we were interested in places of parameter space where {\it adaptation} could be done more efficiently, 
by changing the dynamic state of the system with minimal variation of parameters, 
then our ideal spots in parameter space would be places where the gradient (in parameter space) of $\rho$ is large. 
As an example, in figure \ref{figGrad} we show the modulus and direction of $\nabla\rho$ for the logistic map. 
We have estimated $\nabla\rho$ from finite differences of parameters values.

From the triangular patterns found for {\it Hieronyma moritziana} like species in the case of phenology (figure \ref{hyronima_scan}) and for chaotic dynamics (figure \ref{fig34} (left))
it is clear that lines of slope $-1$ play a role in parameter space $(E,T_p)$. The relation $T_p=E-1$ is significant in the method introduced by May and Sugihara:
it limits the amount of information that in principle could be found in vectors of the form $(x_i,x_{i-1},x_{i-2},\ldots,x_{i-(E-1)})$ when trying to make a prediction about  
$x_{i+T_p}$. The lines of slope $-1$ are perpendicular to lines of the form $T_p=E-1$, which suggests that $\nabla\rho$ (see figure \ref{figGrad}) is very relevant in this method, 
which also highlights the natural framework it provides for considering {\it adaptation} in Complex Systems, as commented in the previous paragraph.

We have presented results for several types of system-dynamics based on the forecasting method introduced by May and Sugihara \cite{MS90}.
Our results highlight the powerfulness of full plots of $\rho(E,T_p)$ obtained with the nonlinear forecasting method to analyze and characterize the dynamics of Complex Systems,
for {\it scales} as large as forest Ecosystems and as small as Proteins. \\

\noindent
{\bf Data availability:} Data available in Zenodo at https://doi.org/10.5281/zenodo.12610823. Also, for more ecology data: https://doi.org/10.5281/zenodo.7662623 (see reference \cite{EcologySaul1}).\\


\begin{thebibliography}{9}

\bibitem{MS90}
  Sugihara G. and May R. M.,
  Nonlinear forecasting as a way of distinguishing chaos from measurement error in time series,
  {\it Nature} {\bf 344}, 734–741 (1990).
  
\bibitem{brain1}
  Bubic A., von Cramon D. Y. , Schubotz R. I.,
  Prediction, cognition and the brain,
  {\it Frontiers in Human Neuroscience} {\bf 4}, Article 25 (2010).
  
\bibitem{EcologySaul1}
  Flores S., Forister M. L., Sulbaran H., D\'iaz R. , Dyer L. A.,
  Extreme drought disrupts plant phenology: Insights from 35 years of cloud forest data in Venezuela,
  {\it Ecology} {\bf 104}, Issue 5, e4012 (2023).
  
\bibitem{Schuster}
  Schuster H. G. ,
  {\it Deterministic Chaos: An Introduction} (Fourth edition),
  Wiley-VCH, 2005. See Chapter 5, fig. 58.

\bibitem{Gonzalez1}
  Gonz\'alez J. A. , Carvalho L. B. , 
  Analytical Solutions to Multivalued Maps,
  {\it Mod. Phys. Lett. B} {\bf 11}, 521 (1997);
  Gonz\'alez J. A. , Reyes L. I. , Guerrero L. E. ,
  Exact solutions to chaotic and stochastic systems,
  {\it Chaos} {\bf 11}, 1 (2001).
  
\bibitem{ghws}
  Reyes L,  Laroze D.,
  Cellular Automata for excitable media on a Complex Network: 
  The effect of network disorder in the collective dynamics,
  {\it Physica A} {\bf 588}, 126552 (2022).

\bibitem{new}
  Parameter values for the GHWS model in figure \ref{ideal}: $K=6$, $\sigma=2.4$, $z=5$, $N=1000$. 
  After $T=1000$ steps we got the fluctuations in activity from the next $1000$ steps. 
  For these parameter values the   normalized fluctuations  
  $\varsigma=N(\langle F^2\rangle-\langle F\rangle^2)/[\langle F\rangle (1-\langle F\rangle)]$ are $\approx 1$, 
  where $F$  is the activity \cite{ghws}. $\varsigma\approx 1$ is equivalent to say that 
  the dynamics is an ideal gas like one (see \cite{Chandler}, section $3.6$).
  
\bibitem{Chandler}
  Chandler D., 
  {\it Introduction To Modern Statistical Mechanics}, 
  Oxford University Press, 1987.

\bibitem{protein1}
  Gonz\'alez-Paz L. {\it et al.},
  Structural deformability induced in proteins of potential interest associated with COVID-19 by binding of homologues present in
  ivermectin: Comparative study based in elastic networks models,  {\it Journal of Molecular Liquids} {\bf 340}, 117284 (2021).
  
\bibitem{protein2}
  Haliloglu T., Bahar I. , Erman B.,
  Gaussian Dynamics of Folded Proteins,
  {\it Physical Review Letters} {\bf 79}, Number 16, 3090 (1997);
  Bahar I., Atilgan A. R., Erman B.,
  Direct evaluation of thermal fluctuations in proteins using a single-parameter harmonic potential,
  {\it Folding \& Design} {\bf 2} (3), 173–181 (1997).
  
\bibitem{moldyn}
  GROMACS:  https://www.gromacs.org/ \\
  
  WebGRO for Macromolecular Simulations: https://simlab.uams.edu/
  
\bibitem{alpha}
  Mehra S., Sahay S., Maji S. K.,
   $\alpha$-Synuclein misfolding and aggregation: Implications in Parkinson's disease pathogenesis, 
   {\it Biochim Biophys Acta Proteins Proteom} {\bf 1867} (10), 890-908 (2019).

\bibitem{alpha2}
  Alza N. P., Iglesias P. A., Conde M. A., Uranga R. M., Salvador G. A.,
  Lipids at the Crossroad of a-Synuclein Function and Dysfunction: Biological and Pathological Implications, 
  {\it Frontiers in Cellular Neuroscience} {\bf 13}, 175  (2019).
  
\bibitem{alpha3}
  Ray S. {\it et al.}, 
  $\alpha$-Synuclein aggregation nucleates through liquid–liquid phase separation, {\it Nature Chemistry} {\bf 12}, 705 – 716 (2020).

\bibitem{alphaN}
   Fusco G. {\it et al.}, 
   Direct observation of the three regions in a-synuclein that determine its membrane-bound behaviour, 
   {\it Nature Comunications} {\bf 5}, 3827 (2014).
  
\end{thebibliography}
\end{document}